\documentclass[twocolumn, aps, prl]{revtex4}
\usepackage{graphicx}
\usepackage{amssymb}
\usepackage{amsmath}

\def\lsim{\lower -0.3ex \hbox{$<$} \kern -0.75em \lower 0.7ex \hbox{$\sim$}}
\def\gsim{\lower -0.3ex \hbox{$>$} \kern -0.75em \lower 0.7ex \hbox{$\sim$}}

\def\Vec#1{{\bf #1}}

\def\vare{\varepsilon}

\begin{document}


\title{Electron delocalization 
in bilayer graphene induced by an electric field}

\author{Mikito Koshino}

\affiliation{
Department of Physics, Tokyo Institute of Technology, 
Meguro-ku, Tokyo 152-8551, Japan}
\date{\today}

\begin{abstract}
Electronic localization is numerically studied
in disordered bilayer graphene with an electric-field induced
energy gap.
Bilayer graphene is a zero-gap semiconductor,
in which an energy gap can be opened and controlled
by an external electric field perpendicular to the layer plane.
We found that, in the smooth disorder potential not mixing 
the states in different valleys ($K$ and $K'$ points),
the gap opening causes a phase transition
at which the electronic localization length diverges.
We show that this can be interpreted as 
the integer quantum Hall transition at each single valley,
even though the magnetic field is absent.
\end{abstract}

\maketitle


Since the experimental discovery 
of monatomic graphene \cite{Novo04,Novo05,Zhan05-2},
the electronic properties of graphene-related materials
have been extensively studied.
The graphene bilayer, which is also experimentally 
avaliable \cite{Novo05,Novo06,Ohta06}, was shown to have a 
unique band structure distinct from monolayer,
where the conduction and valence bands with quadratic dispersion
touch at $K$ and $K'$ points in the Brillouin zone \cite{McCa06a}.
The transport properties of the bilayer graphene
have been studied experimentally, \cite{Novo06,Gorb07,Moro08}
and also extensively studied in theories
\cite{Kosh06,Katz06,Snym07,Cser07a,Cser07b}.
The quantum correction to the conductivity,
which is important in the low temperature,
was studied for monolayer graphene \cite{Suzu02,Moro06}
and for bilayer. \cite{Kech07,Gorb07}
A unique property of bilayer graphene
is that an electric-field applied perpendicularly 
to the layers opens an energy gap
between the electron and hole bands
\cite{McCa06a,McCa06b,Cast07,Nils07a,Hong07}.
The transport property of the gapped bilayer graphene
was also studied. \cite{Nils07b}
The electric-field induced energy gap was observed
in recent experiments. \cite{Ohta06,Oost07}

When the impurity potential is smooth compared to the atomic scale
and the intervalley scattering is negligible,
we can treat two valleys (around $K$ or $K'$) as independent subsystems.
The sub-Hamiltonian within each valley 
generally has no time-reversal symmetry in itself,
since the time-reversal counterpart of the states at $K$  
exists at $K'$.  \cite{Suzu02,Kech07}
In bilayer graphene with a gap, interestingly, 
each sub-Hamiltonian 
has non-zero Hall conductivity even in zero magnetic field.
The monolayer graphene can also have an energy gap
and the corresponding Hall conductivity
when the sub-lattice symmetry is broken \cite{Seme84,Ludw94}, 
while it cannot be controlled externally. \cite{Giov07}

The single-valley Hall conductivity can never be directly observed,
since it is exactly canceled by the contribution from the other valley.
In this paper, however,
we find that this quantity strongly influences the electron localization
properties, and this may be observed in the longitudinal conductivity.
We show that, when the system is 
under a smooth random potential not mixing valleys, 
the perpendicular electric field causes 
a phase transition at which the localization length diverges.
We find that this is interpreted as 
the quantum Hall transition at each single valley,
where the single-valley Hall conductivity changes from one integer to
another.
To demonstrate this, we numerically calculate the localization length
in disordered bilayer graphene with electric fields,
to actually 
show that the localization length diverges at a certain field amplitude.
We then calculate the single-valley Hall conductivity
and show that the delocalization is actually associated with
the transition between different quantum Hall phases.


\begin{figure}
\begin{center}
 \leavevmode\includegraphics[width=50mm]{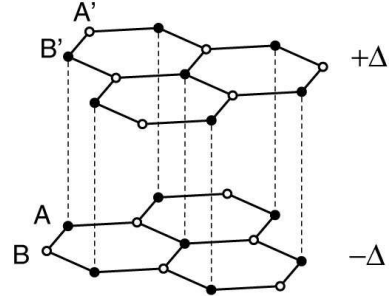}
\end{center}
\caption{Atomic structure of
the bilayer graphene.
$+\Delta$ and $-\Delta$ represents the potential of the top and bottom layers,
respectively.}
\label{fig_atom}
\end{figure}


Bilayer graphene is composed of a pair of hexagonal networks of
carbon atoms, which include $A$ and $B$ atoms on the bottom layer and $A'$
and $B'$ on the top. 
As shown in Fig, \ref{fig_atom},
both layers are arranged in the Bernal stacking,
where $A$ atoms are located directly below $B'$ atoms. \cite{McCa06a}
The low energy spectrum is given by
the states around $K$ and $K'$ points in the Brillouin zone.
Neighboring $A$ and $B'$ sites are coupled to create 
high-energy bands, and remaining $A'$ and $B$ sites form the
low-energy bands touching at the zero energy.
The effective low-energy Hamiltonian around $K$-point reads
\cite{McCa06a}
\begin{equation}
 {\cal H}^K_0 = 
\begin{pmatrix}
\Delta &  \frac{\hbar^2}{2m^*} (k_x - ik_y)^2 \\
\frac{\hbar^2}{2m^*}(k_x + ik_y)^2  & -\Delta,
\end{pmatrix}
\label{eq_H0} 
\end{equation}
which operates on a wave function
$(F^K_{A'},F^K_B)$, where $F^K_{X}$
represents the value of the envelope function of $K$ point at site $X$.
Here $\Delta$ and $-\Delta$
are the external electrostatic potential 
at the top and bottom layers, respectively,
due to the applied electric field,
and $m^*$ is the effective band
mass defined by $m^* = \gamma_1 / (2v^2)$,
where $\gamma_1$ is the coupling parameter 
for the vertical bonds between
$A$ and $B'$ atoms, and $v$ is the 
velocity of the monolayer graphene. \cite{McCa06a}
We assume that $\Delta$ and $vk$ are much smaller than $\gamma_1$
and neglect the terms more than the third order in 
$\Delta/\gamma_1$ and $vk/\gamma_1$.
The band structure of a real bilayer graphene 
is trigonally warped due to the extra coupling parameter
between $A'$ and $B$, \cite{McCa06a, Kosh06}
but is neglected here for simplicity.
The eigen energy of Eq. (\ref{eq_H0}) is given by
\begin{equation}
 \vare_{\Vec{k} s}= s \sqrt{\left(\frac{\hbar^2k^2}{2m^*}\right)^2 + \Delta^2},
\end{equation}
with $s=\pm$ and $k = \sqrt{k_x^2+k_y^2}$. The energy gap 
extends between $\vare = \pm\Delta$.
The effective Hamiltonian for $K'$ can be obtained by 
exchanging $k_x+ik_y$ and $k_x-ik_y$, giving 
the basically same spectrum.
In absence of $\Delta$, the density of states (DOS) becomes constant,
$\rho_0 = m^*/(2\pi\hbar^2)$ per valley and per spin.


For the disorder potential, we assume that 
the length scale is much longer than the atomic scale,
and neglect the intervalley scattering.
This should be valid as long as the phase coherent time is smaller than the 
intervalley scattering time. \cite{Suzu02,Kech07}
We also assume that the disorder length scale is shorter than 
the typical wavelength of $2\pi/k$ with $k$ being the wave number
from $K$ or $K'$ points, so as to be modeled as a 
short-ranged potential within the valley decoupled Hamiltonian.
This is then expressed as \cite{Kosh06}
\begin{equation}
 V = 
\sum_i u_i \delta(\Vec{r} - \Vec{r}_i)
\begin{pmatrix}
1 & 0 \\
0 & 1
\end{pmatrix}.
\label{eq_V}
\end{equation}
We assume an equal amount of positive and negative 
scatterers $u_i = \pm u$ and a total density per unit area $n_{\rm imp}$.
At $\Delta=0$, the energy broadening $\Gamma = \hbar/(2\tau)$ 
becomes independent of the energy
in the weak disorder limit, and is expressed by \cite{Kosh06}
\begin{equation}
 \Gamma = \frac{\pi}{2} n_{\rm imp} u^2 \rho_0.
\end{equation}
This will be used as the energy scale characterizing the 
disorder strength.
We also introduce unit wave number $k_0$ as
$\hbar^2k_0^2/(2m^*) = \Gamma/2$ and unit length
$\lambda_0 = 2\pi/k_0$.

The single-valley Hamiltonian with disordered potential
${\cal H}^K = {\cal H}^K_0 + V$ belongs to the unitary symmetry
class when $\Delta$ is non-zero.
It becomes the orthogonal class only at $\Delta= 0$,
as we have the relation $\sigma_x {\cal H}^K \sigma_x = ({\cal H}^{K})^*$
with the Pauli matrix $\sigma_x$.
This is an effective time-reversal symmetry
within a single valley and should be distinguished from
the real time-reversal symmetry connecting $K$ and $K'$
which always exists at any values of $\Delta$.

The single-valley Hamiltonian
has a non-zero Hall conductivity when $\Delta \neq 0$.
This is estimated by Kubo formula,
\begin{equation}
 \sigma_{xy} = \frac{\hbar e^2}{iS}
\sum_{\alpha,\beta}
\frac{f(\vare_\alpha)-f(\vare_\beta)}{\vare_\alpha-\vare_\beta}
\frac{
\langle\alpha|v_x|\beta\rangle
\langle\beta|v_y|\alpha\rangle
}
{\vare_\alpha-\vare_\beta+i\delta},
\label{eq_kubo_formula}
\end{equation}
where $S$ is the area of the system,
$v_x$ and $v_y$ are the velocity operators, 
$\delta$ is the positive infinitesimal,
$f(\vare)$ is the Fermi distribution function,
and
$| \alpha\rangle$ and $\vare_\alpha$
describe the eigenstate and
the eigen energy of the system.

By applying this formula to the ideal $K$-point 
Hamiltonian ${\cal H}^K_0$ at zero temperature,
we obtain
\begin{equation}
\sigma_{xy}^{K}
= \left\{
\begin{array}{cc}
\displaystyle \frac{e^2}{h} \frac{\Delta}{|\vare_F|}
& \quad |\vare_F| > |\Delta|
\\
\displaystyle \frac{e^2}{h} {\rm sgn}(\Delta)
& \quad |\vare_F| < |\Delta|,
\end{array}
\right.
\label{eq_sigma_xy}
\end{equation}
with ${\rm sgn}(x) = x/|x|$.
The result for $K'$ point is given by 
$\sigma_{xy}^{K'}=-\sigma_{xy}^{K}$, so that the 
net Hall conductivity is exactly zero as it should.
$\sigma_{xy}^{K}$ at the energy gap
is quantized into different values in the negative 
and positive $\Delta$'s.
When $\vare_F$ is fixed to zero and $\Delta$ is 
continuously changed from negative to positive,
we have the Hall conductivity change from
$\sigma^K_{xy} = -e^2/h$ to $e^2/h$,
and $\sigma^{K'}_{xy} = e^2/h$ to $-e^2/h$.

\begin{figure}
\begin{center}
 \leavevmode\includegraphics[width=60mm]{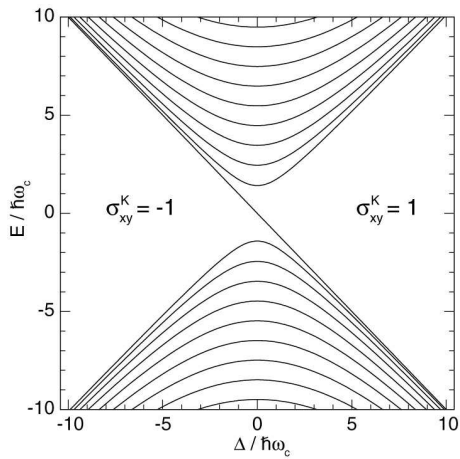}
\end{center}
\caption{Energy spectrum Eq. (\ref{eq_LL})
plotted against $\Delta$,
at $K$ valley in the bilayer graphene 
with a uniform magnetic field.
}
\label{fig_schm}
\end{figure}

The jump of the Hall conductivity 
can be intuitively explained by the consideration of
the energy spectrum in the magnetic field.
In a uniform field $B$, the Landau level energy
at $K$ is obtained by substituting $\Vec{k}$ 
in Eq. (\ref{eq_H0}) with $\Vec{k} + e \Vec{A}/\hbar$ 
with the vector potential $\Vec{A} = (0, Bx)$. \cite{McCa06a}
This results in
\begin{eqnarray}
&& \vare_{s,n}
= s \sqrt{(\hbar\omega_c)^2n(n-1) + \Delta^2} 
\quad (n \geq 2) \nonumber\\
&& \vare_0 =  \vare_1  = -\Delta.
\label{eq_LL}
\end{eqnarray}
where $\omega_c = eB/m^*$ and $s=\pm$ represents 
the electron and hole bands.
The energy spectrum is plotted 
as a function of $\Delta$ in Fig. \ref{fig_schm},
We have doubly-degenerate Landau levels $\vare_0$ and $\vare_1$
at $E = -\Delta$,
which moves from the electron band down to the hole band 
as $\Delta$ increases.
When the Fermi energy is fixed at $E_F=0$ and
$\Delta$ is changed from negative to positive, 
those two Landau levels cross $E_F$, and therefore give a change of 
$\sigma_{xy}^K$ by 
$2e^2/h$, twice as large as a Hall conductivity quantum.
It is important that this discontinuity in $\sigma_{xy}^K$ 
is not dissolved in the limit of $B\rightarrow 0$,
and is actually consistent with the discussion above 
in zero magnetic field.
For $K'$ point, we have $\vare_0 =  \vare_1  = +\Delta$,
leading to the opposite sign of the Hall conductivity change.

Now we numerically calculate the electronic states
in the disorder potential by exactly diagonalizing the
Hamiltonian matrix. 
We consider a finite square system with $L \times L$
described by the $K$-point Hamiltonian
${\cal H}^K = {\cal H}^K_0 + V$,
imposing a boundary condition with phase factors
$\exp(i\phi_x)$ and $\exp(i\phi_y)$ for $x$- and $y$-directions,
respectively.
To make the matrix finite, 
we introduce the $k$-space cut-off $k_c  = 6k_0$, 
which corresponds to $\vare_c = 18\Gamma$.  

To investigate the localization property,
we calculate the Thouless number $g$, which is the 
ratio of the shift $\Delta E$ of each energy level
due to the change of the boundary condition, to the
level spacing $\left[L^2 \rho\right]^{-1}$ with $\rho$ being the density of
states per unit area. \cite{Imry}
We estimate the energy shift by 
$\Delta E = \pi \langle |\partial^2 E(\phi)/\partial \phi^2|\rangle$,
where $E(\phi)$ represents the eigen energy as a function of the 
boundary phase $\phi=\phi_x$ with fixed $\phi_y$, 
and $\langle\,\rangle$ represents
averaging over different levels in a small energy region around 
the energy $\vare$ in question.
The localization length $L_{\rm loc}$ is estimated by fitting the results 
to $g(L) \propto \exp(-L/L_{\rm loc})$.
When $g \ll 1$, $g$ is approximately related
to the longitudinal conductivity $\sigma_{xx}$ by
$\sigma_{xx} \sim (e^2/\hbar) g.$ \cite{Imry}
We also calculate the Hall conductivity
by substituting the eigenstates of the disordered system
to Kubo formula, Eq. (\ref{eq_kubo_formula}).
For every quantity, we take an average over a number of samples
with different configuration of the disorder potential
and boundary phase factors $\phi_x, \phi_y$.


The absolute value of $\sigma_{xy}$ generally depends on 
the $k$-space cut-off. This is because $\sigma_{xy}$ can be expressed 
as the summation of the contribution from all the occupied states
below the Fermi energy, 
unlike $\sigma_{xx}$ which only
depends on the states at the Fermi energy. 
In the clean limit, the cut-off at $\vare = \pm\vare_c$
leads to overall shift of $\sigma_{xy}(\vare)$ 
by a constant, $(e^2/h)\Delta/\vare_c$,
which is the contribution from the missing states out of the cut-off.
$\sigma_{xy}$ in the disordered system has a similar shift
which vanishes in the limit of $\vare_c \rightarrow 0$.
However, as long as $\vare_c$ is much larger
than the energy broadening $\Gamma$,
it hardly changes
the function form of $\sigma_{xy}(\vare)$ around $\vare= 0$
except for this small constant.

\begin{figure}
\begin{center}
 \leavevmode\includegraphics[width=80mm]{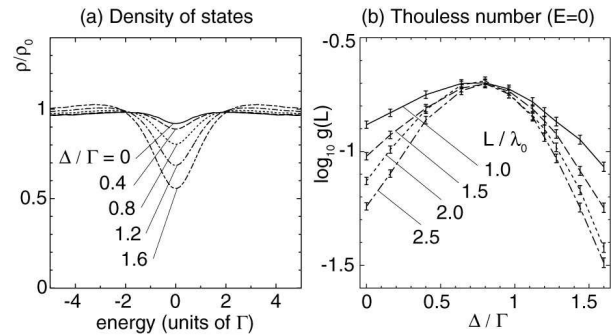}
\end{center}
\caption{
(a) Density of states as a function of energy,
in bilayer graphene with several $\Delta$'s at fixed $\Gamma$.
(b) Thouless number $g$ at zero energy plotted against $\Delta$.
}
\label{fig_dos_thou}
\end{figure}
\begin{figure}
\begin{center}
 \leavevmode\includegraphics[width=60mm]{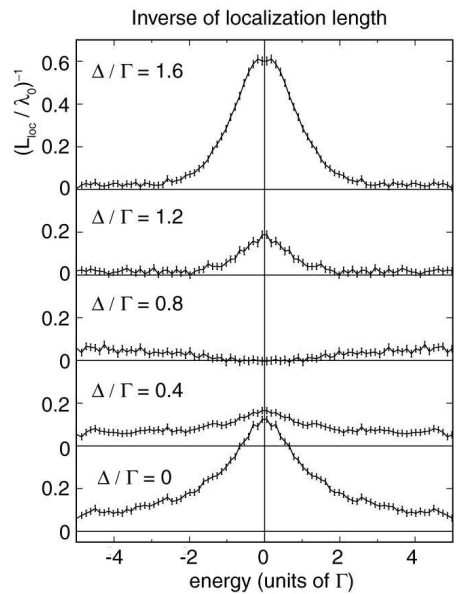}
\end{center}
\caption{
Inverse of the localization as a function of energy
in bilayer graphene with several $\Delta$'s.
}
\label{fig_loc}
\end{figure}

\begin{figure}
\begin{center}
\leavevmode\includegraphics[width=70mm]{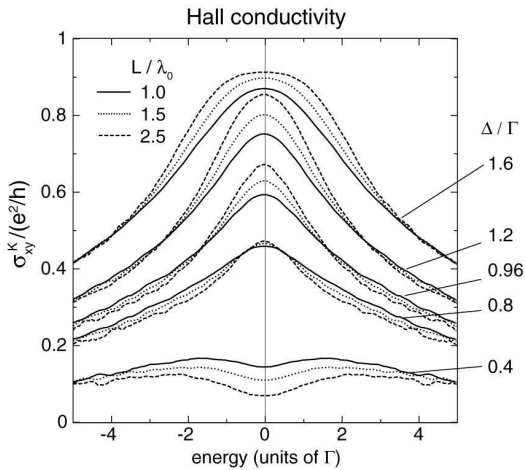}
\end{center}
\caption{
Hall conductivity 
of a single-valley ($K$) Hamiltonian of bilayer graphene
with several $\Delta$'s, plotted against the energy.
}
\label{fig_hall}
\end{figure}

\begin{figure}
\begin{center}
 \leavevmode\includegraphics[width=60mm]{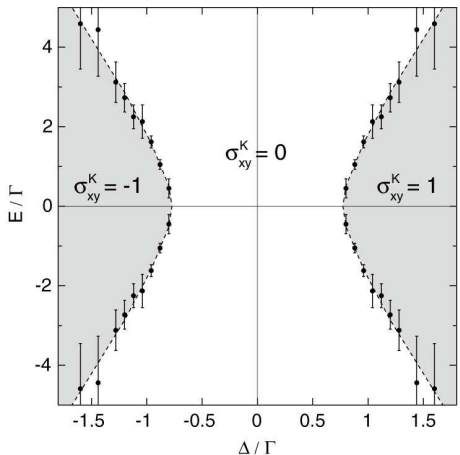}
\end{center}
\caption{
Hall plateau diagram against $\Delta$ and energy $E$.
Dots represent the critical energies
estimated from the size-dependence of $\sigma^K_{xy}$.
Dashed curves are guide for eyes for the phase boundary.
}
\label{fig_phase}
\end{figure}


Figure \ref{fig_dos_thou} (a) shows the DOS
in disordered bilayer graphene, with various $\Delta$'s at fixed $\Gamma$.
The energy axis is scaled in units of $\Gamma$.
As $\Delta$ increases, the DOS at the zero energy
monotonically decreases as expected,
while the disorder potential fills the energy gap with some states.
In Fig. \ref{fig_dos_thou} (b), we plot the Thouless number $g$
at zero energy
as a function of $\Delta$, for several system sizes.
We naively expect that the conductivity at zero energy
monotonically decreases in increasing $\Delta$
since the DOS becomes smaller,
but this is not the case here.
At every system size, $g$ goes up first when $\Delta$
starts from 0, and then changes to decrease at 
$\Delta \sim 0.8\Gamma$.
Significantly $g$ becomes independent of the system size 
around the peak, while in other region it drops exponentially
as the system size increases.
This suggests that the electronic state at zero energy
is localized both in small and large $\Delta$'s,
but delocalized only at $\Delta \sim 0.8\Gamma$.
We calculate $g$ also for other energies
and obtain the localization length $L_{\rm loc}$ 
as a function of energy from its size dependence.
In Fig. \ref{fig_loc}, we plot $1/L_{\rm loc}$ against energy 
for several values of $\Delta$.
As $\Delta$ increases from zero, $1/L_{\rm loc}$ gradually decreases,
and the value around $\vare=0$ reaches almost zero 
(i.e., the states extended) at $\Delta \sim 0.8\Gamma$.
The center value recovers to non-zero again in even larger $\Delta$, 
and the peak grows rapidly as $\Delta$ increases; the states 
at the gap region become localized more and more.

The corresponding plots for the single-valley
Hall conductivity $\sigma^K_{xy}$ are shown in Fig. \ref{fig_hall}.
$\sigma^K_{xy}$ becomes completely 0 at $\Delta=0$ (not shown),
and the value generally goes up as $\Delta$ increases.
At each single $\Delta$, we notice that 
$\sigma^K_{xy}$ moves in a specific direction
as the system size $L$ expands.
At $\Delta =0.4\Gamma$, $\sigma^K_{xy}$ 
decreases in all the energy as $L$ increases.
At $\Delta =1.6\Gamma$, on the contrary,
$\sigma^K_{xy}$ around $\vare =0$
evolves in the upward direction, oppositely.
The critical value of $\Delta$ where the 
$\sigma^K_{xy}$ switches its evolving direction
depends on the energy.
At $E=0$, this is close to $\Delta/\Gamma = 0.8$.

In the integer quantum Hall effect,
the Hall conductivity in the infinite system
is quantized as long as the states at the Fermi energy 
are localized. \cite{Prui}
In finite-size systems, the Hall conductivity 
averaged over disorder configurations
generally takes non-integer values,
while it gradually approaches 
the quantized value as the system size expands. \cite{Ando86a}
Therefore the quantized Hall conductivity 
in the infinite system can be estimated
from the scaling behavior of finite-size values. 
In above results, we expect that the region of decreasing $\sigma^K_{xy}$
becomes the Hall plateau with $\sigma^K_{xy}=0$,
and that of increasing $\sigma^K_{xy}$ becomes the plateau with
$e^2/h$, in an infinite system size.
The point where the $\sigma^K_{xy}$ switches the scaling dependence
can be regarded as the phase transition between
the different Hall phases $\sigma^K_{xy}=0$ and 1. 
Significantly this point 
coincides well with the divergence of the localization 
length $L_{\rm loc}$ in Fig. \ref{fig_loc}.
The feature is most striking at $\Delta= 0.8\Gamma$,
where $\sigma^K_{xy}$ at $\vare = 0$ is about to change its size dependence, 
and $L_{\rm loc}$ at the center is also near divergence.
This is a consistent result since the electronic states must be
delocalized at the critical point separating different Hall plateaus.
\cite{Prui}

Indeed, when the energy gap width 
is much larger than the disorder strength, i.e., $\Delta/\Gamma \gg 1$,
we reasonably expect that 
$\sigma^K_{xy}$ around zero energy is quantized 
at $e^2/h$, the mid-gap value in the clean-limit,
since the gap remains almost intact in the weak disorder.
When the energy gap is small enough that $\Delta/\Gamma \ll 1$,
on the other hand, it is natural that $\sigma^K_{xy}$ vanishes to
zero as in the zero-gap case, since the effect of $\Delta$ is completely 
washed out by the strong disorder.
It is then inevitable to have the phase transition
at the intermediate value of $\Delta/\Gamma$, which is
presumably of the order of 1,
while the factor may depend on the specific disorder model.

At $\Delta > 0.8$,
we have two critical energies 
which separates upward and downward moving region in $\sigma^K_{xy}$
in Fig. \ref{fig_hall}.
The localization length in Fig. \ref{fig_loc}
is huge around the corresponding energies,
so that it becomes harder to specify the critical points out of them
due to the numerical error.
Fig. \ref{fig_phase} shows
the phase diagram speculated in the infinite system,
where each phase corresponds to the Hall plateau with 
quantized $\sigma^K_{xy}$.
We determine the phase boundary by taking points
where $\sigma^K_{xy}$ changes its moving direction.
The dashed curves separating the phases are guides for eyes.
The phase diagram is symmetric with respect to $\Delta=0$,
while the sign of the Hall conductivity is opposite 
between negative and positive $\Delta$'s.


We cannot eliminate the delocalized states
as long as $K$ and $K'$ valleys are decoupled,
in the following sense:
In the clean system, as previously discussed,
the single-valley Hall conductivity $\sigma_{xy}^K$
at the gap is quantized at $-1$ and 1 (in units of $e^2/h$) 
at negative and positive $\Delta$, respectively.
In the disordered system,
we expect that $\sigma_{xy}^K$ at the gap
remains quantized when the gap width is large enough
compared to the energy broadening due to the disorder.
Thus the Hall conductivity at a fixed Fermi energy
definitely changes from $-1$ to 1
when $\Delta$ changes from $-\Delta_0$ to $+ \Delta_0$
with sufficiently large $\Delta_0$.
This requires that the delocalized states appear
at that energy somewhere between $-\Delta_0$ and $+ \Delta_0$,
because otherwise the Hall conductivity stays constant. \cite{Prui}
The present calculation indeed suggests that
there are two delocalized points of $\Delta$ (negative and positive)
at each fixed energy.

The delocalized states  
would disappear in presence of 
the short-ranged disorder with atomic length scale,
which induces intervalley coupling between $K$ and $K'$
and makes the original time-reversal symmetry effective.
The localization length at the phase boundary
is expected to diverge as the inter-valley scattering rate goes to zero.
The trigonal warping effect due to the extra hopping parameter, 
which is also neglected here, does not kill the delocalized states 
since it does not mix the valleys, while it generally weaken
the electron localization at $\Delta = 0$. \cite{Kech07}


In this paper we proposed that the external electric field
applied to the bilayer graphene
causes a phase transition
at which the localization length diverges.
The transition can be interpreted as
an analog of quantum Hall transition at each single valley.
While the actual Hall conductivity is never directly observed, 
it may be possible to observe
the divergence of the localization length
in a temperature ($T$) dependence of the conductivity $\sigma_{xx}$, 
as in the conventional integer quantum Hall effect. \cite{Wei94}
Figure \ref{fig_dos_thou}(b) can be approximately viewed as 
plots of $\sigma_{xx}$ at different $T$'s,
when we use a relation $\sigma_{xx} \sim (e^2/\hbar) g$,
and regard $L$ as the phase coherent length at $T$.
When the system goes through the phase transition,
$\sigma_{xx}$ exhibits a peak at the phase boundary
and its peak width  becomes narrower as the temperature is lowered. 
\cite{Wei94}
To observe this, the temperature must be
low enough that the phase coherent length exceeds
the typical localization length $l_B$. 
Our calculation is based on the a typical disorder model used in theories,
while it may not describe some specific situations
in the real bilayer graphene currently available,
which is supposed to include puddles (long-range inhomogeneity)
or impurity-bound states described by the variable-range hopping model.
We leave to the future work 
the study on those effects on the present phenomenon.



The author acknowledges helpful interactions with K. Nomura,
S. Ryu, B. Altshuler, I. L. Aleiner, B. \"{O}zyilmaz and P. Kim.
This work has been supported in part by the 21st Century COE Program at
Tokyo Tech \lq\lq Nanometer-Scale Quantum Physics'' and by Grants-in-Aid
for Scientific Research from the Ministry of Education, 
Culture, Sports, Science and Technology, Japan.



\begin{thebibliography}{99}


\bibitem{Novo04}
K. S. Novoselov 
A. K. Geim, S. V. Morozov, D. Jiang, Y. Zhang, S. V. Dubonos, 
I. V. Grigorieva, and A. A. Firsov, 
Science {\bf 306}, 666 (2004). 

\bibitem{Novo05}
K. S. Novoselov, 
A. K. Geim, S. V. Morozov, D. Jiang, M. I. Katsnelson, 
I. V. Grigorieva, S. V. Dubonos, and A. A. Firsov, 
Nature {\bf 438}, 197 (2005).
 
\bibitem{Zhan05-2}
Y. Zhang, 
Y. W. Tan, H. L. Stormer, and P. Kim, 
Nature {\bf 438}, 201 (2005).

\bibitem{Novo06}
K. S. Novoselov, 
E. McCann, S. V. Morozov, V. I. Fa'lko,
M. I. Katsnelson, U. Zeitler, D. Jiang, F. Schedin, and A. K. Geim, 
Nat. Phys. {\bf 2}, 177 (2006).

\bibitem{Ohta06}
T. Ohta, 
A. Bostwick, T. Seyller, K. Horn, and E. Rotenberg, 
Science {\bf 313}, 951 (2006).


%
\bibitem{McCa06a}
E. McCann and V. I. Fal'ko, Phys. Rev. Lett. {\bf 96}, 086805 (2006).

%
\bibitem{Gorb07}
R.V. Gorbachev, 
F.V. Tikhonenko, A. S. Mayorov, D.W. Horsell, and A. K. Savchenko
Phys. Rev. Lett. {\bf 98}, 176805 (2007).

\bibitem{Moro08}
S.V. Morozov, 
K. S. Novoselov, M. I. Katsnelson,
F. Schedin, D. C. Elias, J. A. Jaszczak, and A. K. Geim,
Phys. Rev. Lett. {\bf 100}, 016602 (2008).



\bibitem{Kosh06}
M. Koshino and T. Ando, Phys.\ Rev.\ B {\bf 73}, 245403 (2006).

\bibitem{Katz06}
M. I. Katsnelson, Eur. Phys. J. B {\bf 52}, 151 (2006).

\bibitem{Snym07}
I. Snyman and C. W. J. Beenakker, Phys. Rev. B {\bf 75}, 045322 (2007).

\bibitem{Cser07a}
J. Cserti, Phys. Rev. B {\bf 75}, 033405 (2007).

\bibitem{Cser07b}
J. Cserti, 
A. Csordas, and G. David,
Phys. Rev. Lett. {\bf 99}, 066802 (2007)

\bibitem{Kech07}
K. Kechedzhi, V. I. Fal'ko, E. McCann, and B. L. Altshuler,
Phys. Rev. Lett. {\bf 98}, 176806 (2007).

\bibitem{Suzu02} 
H. Suzuura and T. Ando, 
Phys.\ Rev.\ Lett.\ {\bf 89}, 266603 (2002); 
E. McCann, K. Kechedzhi, V. I. Falko, H. Suzuura, T. Ando, and B.
L. Altshuler, \textit{ibid}.\ {\bf 97}, 146805 (2006); 
I. L. Aleiner and K. B. Efetov, \textit{ibid}. {\bf 97}, 236801 (2006).
A. Altland, \textit{ibid}.\ \textbf{97}, 236802 (2006);
K. Nomura, M. Koshino, and S. Ryu, \textit{ibid}. {\bf 99}, 146806 (2007).

\bibitem{Moro06}
S. V. Morozov, K. S. Novoselov, M. I. Katsnelson, F. Schedin, L.
A. Ponomarenko, D. Jiang, and A. K. Geim, 
Phys. Rev. Lett. {\bf 97}, 016801 (2006),
X. Wu, X. Li, Z. Song, C. Berger, and W. A. de Heer,
\textit{ibid}. {\bf 98}, 136801 (2007),
H. B. Heersche, P. Jarillo-Herrero, 
J. B. Oostinga, L. M. K. Vandersypen, and A. F. Morpurgo,
Nature (London) {\bf 446}, 56 (2007).


\bibitem{McCa06b} 
E. McCann Phys. Rev. B {\bf 74}, 161403(R) (2006).

\bibitem{Cast07} 
E. V. Castro, 
K. S. Novoselov, S.V. Morozov, N. M. R. Peres, J. M. B. Lopes dos Santos, Johan Nilsson, F. Guinea, A. K. Geim, and A. H. Castro Neto,
Phys. Rev. Lett. {\bf 99}, 216802 (2007).

\bibitem{Hong07}
Hongki Min,  
Bhagawan Sahu, Sanjay K. Banerjee, and A. H. MacDonald,
Phys. Rev. B {\bf 75}, 155115 (2007).

\bibitem{Nils07a}
J. Nilsson and A. H. Castro Neto, Phys. Rev. Lett. {\bf 98}, 126801 (2007).

\bibitem{Nils07b}
J. Nilsson, 
A. H. Castro Neto, F. Guinea, and N. M. R. Peres,
Phys. Rev. B {\bf 76}, 165416 (2007).



\bibitem{Oost07}
J. B. Oostinga, 
H. B. Heersche, X. Liu, A. F. Morpurgo and L. M. K. Vandersypen,
Nature Mater. {\bf 7}, 151 (2008).


%
\bibitem{Giov07}
G. Giovannetti, 
P. A. Khomyakov, G. Brocks, P. J. Kelly, and J.van den Brink, 
Phys. Rev. B {\bf 76}, 073103 (2007).


%
\bibitem{Imry}
Y. Imry, 
{\it Introduction to Mesoscopic Physics}, 
Oxford Univ. Press, New York and Oxford, 1997, and references therein.

%
\bibitem{Seme84}
G. W. Semenoff, Phys. Rev. Lett. {\bf 53}, 2449 (1984).

%
\bibitem{Ludw94}
A.\ W.\ W.\ Ludwig, M.\ P.\ A.\ Fisher, R.\ Shankar, and G.\ Grinstein, 
Phys.\ Rev.\ B \textbf{50}, 7526 (1994).

%
\bibitem{Prui}
A. M. M. Pruisken: {\it The Quantum Hall Effect},
eds. R. E. Prange and S. M. Girvin (Springer-Verlag, Berlin, 1990).

%
\bibitem{Ando86a}
T. Ando, J. Phys. Soc. Jpn. {\bf 55}, 3199 (1986).

%
\bibitem{Wei94}
H. P. Wei, 
D. C. Tsui, M. A. Paalanen, and A. M. M. Pruisken,
Phys. Rev. Lett. {\bf 61}, 1294 (1988).

%

\end{thebibliography}
\end{document}